\documentclass[10pt,a4paper]{article}
\usepackage{amsmath}
\usepackage{amsfonts}
\usepackage{amssymb}
\usepackage{graphicx}
\usepackage{textcomp}
\usepackage{url}
\usepackage{lipsum}
\usepackage{setspace}
\usepackage{color}
\usepackage{comment}
\usepackage{multirow}
\usepackage[numbers,sort&compress]{natbib}
\usepackage[usenames,dvipsnames,svgnames,table]{xcolor}
\usepackage[lofdepth,lotdepth,caption=false]{subfig}
\usepackage[pdfstartview=FitH]{hyperref}
\usepackage{hyperref}
\newsavebox{\bigleftbox}
\hypersetup{
 colorlinks,
 citecolor=Blue,
 linkcolor=Blue,
 urlcolor=Blue
}

\makeatletter
 \def\footnoterule{\kern-3\p@
   \noindent\hrulefill \kern 2.8\p@} 
\makeatother
\usepackage[left=3.00cm, right=3.00cm, top=2.00cm, bottom=2.00cm]{geometry}

\usepackage{fancyhdr}

\title{\textbf{A DFT Study on the Mechanical, Electronic, Thermodynamic, and Optical Properties of GaN and AlN Counterparts of Biphenylene Network}}
\author{
	Kleuton Antunes Lopes Lima$^{\dag}$ and
	Luiz Antonio Ribeiro Junior$^{\S}$
	}
\date{}

\begin{document}
    \maketitle
	\vspace{-0.6cm}
	\begin{center}\small
		\textit{Institute of Physics, University of Bras\'ilia, 70910-900, Bras\'ilia, Brazil}\\
            \textit{Computational Materials Laboratory, LCCMat, Institute of Physics, University of Bras\'ilia, 70910-900, Bras\'ilia, Brazil}\\
		\phantom{.}\\ \hfill
		$^{\dag}$\url{kleutonantuneslopes@prp.uespi.br}
		$^{\S}$\url{ribeirojr@unb.br}\hfill
		\phantom{.}
	\end{center}
	

\onehalfspace

\noindent\textbf{Abstract:}
The biphenylene network (BPN) is notable in recent fabrication efforts to design new 2D materials. The stability of its boron nitride counterpart, BN-BPN, has been confirmed through numerical investigations. In this study, we conducted a density functional theory (DFT) analysis to examine the mechanical, electronic, thermodynamic, and optical properties of two other group-III counterparts of BPN: gallium nitride (BPN-GaN) and aluminum nitride (BPN-AlN). Our findings reveal that the band gap values for BPN-GaN and BPN-AlN are 2.3 eV and 3.2 eV, respectively, at the HSE06 level. Phonon- and energy-formation calculations and ab initio molecular dynamics (AIMD) simulations suggest that BPN-(Al,Ga)N has good structural and dynamic stabilities. BPN-GaN displayed negative phonon frequencies. However, results from AIMD simulations point to its structural integrity with no bond reconstructions at 1000 K. These materials exhibit noteworthy UV activity, promising prospects as UV collectors. The thermodynamic properties reveal that the heat capacity of both BPN-AlN and BPN-GaN increases with temperature, eventually reaching the Dulong-Petit limit at around 800 K. We also performed calculations to determine the electron- and hole-effective masses, charge carrier mobility, elastic stiffness constants, Young's modulus, and Poisson ratio for both BPN-GaN and BPN-AlN, providing valuable insight into their electronic and mechanical properties.

\section{Introduction}

The rapid advancement of nanoelectronics has spurred the exploration of novel materials with unique properties to meet the evolving demands of high-performance electronic devices \cite{osada2012two,liu2017flexible}. In this context, free-standing 2D hexagonal aluminum nitride (AlN) and gallium nitride (GaN) monolayers \cite{hickman2021next,chen2018growth} have reemerged as promising candidates, drawing inspiration from the advent of graphene \cite{geim2009graphene}. They share a hexagonal lattice structure with graphene and offer distinct features and exciting prospects for various optoelectronic applications \cite{ahangari2017interlayer}. One of the critical distinctions lies in the semiconducting band gap properties displayed by AlN and GaN monolayers, making them highly attractive for these applications \cite{wang2021two}. Their wide band gap enables efficient control of charge carriers, rendering them suitable for applications such as transistors \cite{li2023scaling} and light-emitting diodes (LEDs) \cite{yang2014high}. On the contrary, the zero bandgap of graphene limits its usage in devices requiring distinct on/off switching behavior \cite{lee2015chemically}.

The electronic properties of AlN and GaN monolayers arise from their unique atomic constituents, which include aluminum and gallium atoms. These elements exhibit distinct energy levels and bonding characteristics compared to carbon atoms found in graphene \cite{wang2021two}. As a result, AlN and GaN monolayers display diverse band gap energies and optical properties, giving rise to distinctive absorption and emission spectra that differ from those of graphene \cite{abdullah2022electronic,ahangari2017interlayer}. Beyond their electronic characteristics, AlN and GaN monolayers demonstrate exceptional thermal stability and mechanical properties, making them suitable for deployment in high-temperature environments and high-performance devices \cite{smith1996microstructure,gurbuz2017single}. Their robustness and durability make them ideal for electronic applications that require resistance to extreme conditions \cite{bouguen2009high}.

One prominent application of these materials lies in optoelectronics, specifically in developing LEDs \cite{wu2018exfoliation,ryu2022growth}. GaN-based LEDs have significantly impacted the lighting industry by providing long-lasting energy-efficient alternatives to conventional incandescent and fluorescent lighting sources \cite{zhu2016solid}. Group-III nitrides, including AlN and GaN, are also useful in laser diodes \cite{ohta2010future}, power electronic devices \cite{mohammad1996progress,dobrinsky2013iii}, and sensors \cite{schalwig2001group}. The versatility and wide-ranging applications of AlN and GaN monolayers highlight their significance in driving technological advances in various fields, particularly in optoelectronics and solid-state lighting \cite{mohammad1996progress}. Ongoing research and development efforts continue to explore novel materials synthesis techniques and device architectures to unlock their full potential in the ever-evolving landscape of advanced materials and electronic devices \cite{alberi20182019}.

In recent in-silico studies, a series of 2D materials based on group-III elements have emerged as promising candidates \cite{oba2018design,paul2017computational}. These materials share a comparable structure with certain carbon-based monolayers that have already been successfully synthesized \cite{lin2019synthesis,toh2020synthesis,fan2021biphenylene}. One notable example is the boron nitride form of the biphenylene network (BPN) \cite{fan2021biphenylene}, known as BN-BPN \cite{ma2022bn,shahrokhi2017new,rublev2023overlapping,D2CP05995A}, which has been proposed and its stability confirmed in several numerical studies based on distinct computational protocols. However, to fully comprehend the potential of this specific lattice topology and explore its practical applications, it is crucial to characterize other group-III counterparts of BPN. By investigating these counterparts, we can expand our understanding of their properties and identify potential applications for these materials.

Using computational approaches for material design has become essential in research for several compelling reasons. Usually, theoretical predictions of new nanostructured materials preceded their synthesis by years or decades. Categorical examples are the BPN and the 2D monolayer of C$_{60}$, which had the first theoretical works published by Mortazavi in 2017 \cite{rahaman2017metamorphosis}, and Tomanek in 1990 \cite{belavin2000stability}, respectively. Simulating and predicting material properties can be performed by employing computational models, allowing virtual experiments and expediting the discovery and development of new materials. Computational techniques enable researchers to delve into the underlying physics and chemistry of materials, providing insights into their atomic-scale structures, electronic properties, and interactions. This deep understanding helps to elucidate the fundamental mechanisms that govern the behavior of materials and guides the design of materials with tailored properties.

In this study, we employ density functional theory (DFT) calculations to investigate the mechanical, electronic, thermodynamic, and optical properties of two additional group-III counterparts of BPN: gallium nitride (BPN-GaN) and aluminum nitride (BPN-AlN). We gain valuable insights into these intriguing materials' electronic and structural features using our computational approach. One crucial aspect that we focused on was their structural stability. Interestingly, the BPN topology has proven suitable for designing stable group-III nitride materials.

\section{Methodology}

\begin{figure}[htb!]
    \centering
    \includegraphics[width=0.8\linewidth]{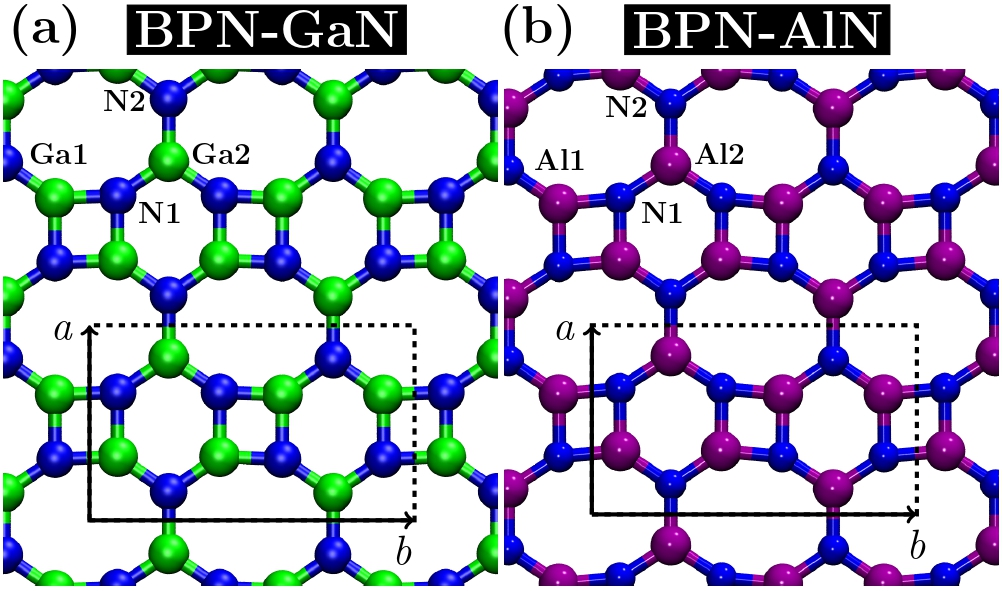}
    \caption{Schematic representation of the monolayers for both (a) BPN-GaN and (b) BPN-AlN. The unit cells are highlighted by rectangles defined by lattice vectors $a$ and $b$. Green, pink, and blue represent gallium, aluminum, and nitrogen atoms.}
    \label{fig:sys}
\end{figure}

The CASTEP \cite{clark2005first} code was used to perform DFT simulations, investigating the mechanical, electronic, thermodynamic, and optical properties of BPN-AlN and BPN-GaN (refer to Figure \ref{fig:sys}). In these simulations, the exchange and correlation functions were treated using the generalized gradient approximation (GGA), parameterized by the Perdew-Burke-Ernzerhof (PBE) functional \cite{perdew1996generalized}, as well as the Heyd-Scuseria-Ernzerhof (HSE06) hybrid functional \cite{heyd2003hybrid}. To accurately account for the interactions between the nuclear electrons of each atomic species, we employed norm-conserving pseudopotentials (as implemented in CASTEP) designed explicitly for gallium, aluminum, and nitrogen.

For achieving electronic self-consistency, we employed the Broyden–Fletcher–Goldfarb–Shannon (BFGS) unrestricted nonlinear iterative algorithm \cite{head1985broyden,PFROMMER1997233}. The plane-wave basis set had an energy cutoff of 700 eV, and we set the convergence criteria for the energy at 1.0 $\times 10^{-5}$ eV. To fully relax the BPN-(Ga, Al)N lattices, we used periodic boundary conditions, ensuring that the residual force on each atom was less than 1.0 $\times 10^{-3}$ eV/\r{A} and the pressure was below 0.01 GPa. During the lattice optimization, the basis vector along the z-direction was held fixed, and we employed a k-grid of $10\times10\times1$, while for electronic and optical calculations, we used a k-grid of $15\times15\times1$ and $5\times5\times1$ for the GGA/PBE and HSE06 methods, respectively. The partial density of states (PDOS) was calculated at the HSE06 level using a k-grid of $20\times20\times1$. The elastic properties were calculated using the LDA/CA-PZ method \cite{PhysRevLett.45.566,PhysRevB.23.5048}. A vacuum region of $20$ \r{A} was considered to avoid spurious interactions among periodic images.

The decision to employ a smaller k-grid resolution in our HSE06 calculations, in contrast to the k-grid used in the GGA/PBE method, is primarily driven by computational considerations, specifically the substantial computational cost associated with the HSE06 method. HSE06-based calculations involve significant computational time, which can limit the feasibility of employing a finer k-grid. We performed convergence tests to validate the suitability of the selected k-grid resolution for our HSE06 calculations. Our convergence tests verified that the chosen k-grid resolution maintains the required accuracy and reliability of the results within an acceptable margin of error.

For the phonon calculations, we utilized the linear response method with a grid spacing of 0.05~1/\r{A} and a convergence tolerance of 10$^{-5}$~eV/\r{A}$^2$. To assess the mechanical properties of the BPN-(Ga,Al)N lattices, we employed the stress–strain method based on the Voigt–Reuss–Hill approximation \cite{Zuo:gl0256,10.1063/1.1709944}. This method follows Hooker's law, establishing a linear relationship between the Lagrangian strain and the Cauchy stress tensor.

We applied an external electric field of 1.0 V/\r{A} along the x and y directions to investigate the optical properties. Using the complex dielectric constant $\epsilon=\epsilon_1+i\epsilon_2$, where $\epsilon_1$ and $\epsilon_2$ represent the real and imaginary parts of the dielectric constant, respectively, we directly obtained the optical quantities. To calculate the imaginary part of the dielectric constant, we employed Fermi's golden rule, which considers interband optical transitions between the valence (VB) and conduction bands (CB), where

\begin{equation}
\epsilon_2(\omega)=\frac{4\pi^2}{V_\Omega\omega^2}\displaystyle\sum_{i\in \mathrm{VB}, \, j\in \mathrm{CB}}\displaystyle\sum_{k} W_k \, |\rho_{ij}|^2 \, \delta	(\varepsilon_{kj}-\varepsilon_{ki}- \hbar \omega).
\end{equation}

\noindent In the context of optical properties, various parameters play a role. These include the photon frequency ($\omega$), the unit cell volume ($V_\Omega$), the k-point weight ($W_k$) in reciprocal space, and the dipole transition matrix element ($\rho_{ij}$). The relationship between the real ($\epsilon_1$) and imaginary ($\epsilon_2$) parts of the dielectric constant is established through the Kramers-Kronig relation \cite{Boukamp_1995}. Specifically, the real part of the dielectric constant can be expressed as:

\begin{equation}
\epsilon_1(\omega)=1+\frac{1}{\pi}P\displaystyle\int_{0}^{\infty}d\omega'\frac{\omega'\epsilon_2(\omega')}{\omega'^2-\omega^2},
\end{equation}
\noindent where $P$ is the Cauchy principal value. 

Additional coefficients can be derived once the dielectric constant's real and imaginary parts are obtained to characterize the optical properties further. These coefficients include the absorption coefficient ($\alpha$), the optical conductivity ($\sigma$), and the reflectivity ($R$). Their calculation involves using $P$ and can be expressed as follows:

\begin{equation}
\alpha (\omega )=\sqrt{2}\omega\bigg[(\epsilon_1^2(\omega)+\epsilon_2^2(\omega))^{1/2}-\epsilon_1(\omega)\bigg ]^{1/2},
\end{equation}
\begin{equation}
\sigma(\omega)= \frac{1}{4\pi}(\omega\epsilon_2(\omega)),
\end{equation}
and
\begin{equation}
R(\omega)=\bigg [\frac{(\epsilon_1(\omega)+i\epsilon_2(\omega))^{1/2}-1}{(\epsilon_1(\omega)+i\epsilon_2(\omega))^{1/2}+1}\bigg ]^2.
\end{equation}

To understand the thermal effect on the BPN-(Ga,Al)N lattices, we used CASTEP to calculate thermodynamic properties, enthalpy $H(T)$, free energy $F(T)$, entropy $S(T)$, and temperature times the entropy term $S(T) = U - F$, where $U$ is the internal energy of the system and the heat capacity of the materials $C_V(T)$ as a function of temperature. The equations below are based on work by Baroni and coworkers \cite{RevModPhys.73.515}. The temperature dependence of the enthalpy $H(T)$ is given by 
\begin{equation}
\displaystyle H(T)=E_{tot}+E_{zp}+\int\frac{\hslash\omega}{\text{exp}\left(\frac{\hslash\omega}{k_BT}\right)-1}N(\omega)d\omega,    
\end{equation}
\noindent where $E_{tot}$ is the total energy, $E_{zp}$ is the zero-point vibrational energy, $k_B$ is the Boltzmann constant, $\hslash$ is the Planck’s constant, and $N(\omega)$ is the phonon density of states. $E_{zp}$ can be evaluated as:
\begin{equation}
\displaystyle E_{zp}=\frac{1}{2}\int N(\omega)\hslash\omega d\omega.    
\end{equation}
The vibrational contribution to the free energy $F(T)$, is written as
\begin{equation}
\displaystyle F(T)=E_{tot}+E_{zp}+k_BT\int N(\omega)\text{ln}\left[1-\text{exp}\left(-\frac{\hslash\omega}{k_BT}\right)\right]d\omega.    
\end{equation}
The entropy $S(T)$ is expressed as follows:
\begin{equation}
\displaystyle S(T)=k_B\int \frac{\frac{\hslash\omega}{k_BT}}{\text{exp}\left(\frac{\hslash\omega}{k_BT}\right)-1}N(\omega)d\omega-k_B\int N(\omega)\text{ln}\left[1-\text{exp}\left(-\frac{\hslash\omega}{K_BT}\right)\right]d\omega.   
\end{equation}
Heat capacity as a function of temperature $C_V(T)$ is obtained according to the equation
\begin{equation}
\displaystyle C_V(T) = k_B \int \frac{\left( \frac{\hbar \omega}{k_B T} \right)^2\text{exp}\left(\frac{\hbar \omega}{k_B T}\right)}{\left[\text{exp}\left(\frac{\hbar \omega}{k_B T}\right) - 1 \right]^2} N(\omega) d\omega.     
\end{equation}

We also calculate the effective mass ($m^*$) of charge carriers, i.e., electrons and holes. This parameter is fundamental in determining the overall carrier mobility in 2D materials. $m^*$ is calculated by fitting the band dispersion to

\begin{equation}
 \displaystyle m^{*}=\hslash^2\left(\frac{\partial^2E(k)}{\partial k^2}\right)^{-1}.   
 \label{mass}
\end{equation}

\section{Results}

We first present the lattice arrangement of the optimized BPN-(Ga,Al)N sheets, as depicted in Figure \ref{fig:sys}. As expected, the GGA/PBE and HSE06 methods yield comparable optimized parameters for these lattices. Therefore, in this regard, we focus on the results obtained using the HSE06 method, showcasing Figure \ref{fig:sys} and Table \ref{tab:bonds}. Figure \ref{fig:sys} shows the slight distortion of the four- and eight-membered rings in the BPN-(Ga,Al)N lattices compared to the all-carbon BPN structure \cite{D2CP05995A,LuoSciRep} while maintaining a planar configuration. These lattices possess an orthorhombic crystal structure and fall under the PMMA space group (D2H-5). The $a$ and $b$ lattice vectors are 9.71 \r{A} and 5.79 \r{A} for BPN-GaN and 9.61 \r{A} and 5.62 \r{A} for BPN-AlN.

\begin{table}[htb!]
\caption{Bond distances for the atoms highlighted in Figure \ref{fig:sys} calculated at the HSE06 level.}
\centering
\label{tab:bonds}
\begin{tabular}{|c|c|c|c|}
\hline
Bond Type & Bond Distance (\r{A}) & Bond Type & Bond Distance (\r{A}) \\ \hline
Ga1-N1    & 1.85                          & Al1-N1    & 1.81                          \\ \hline
Ga2-N1    & 1.82                          & Al2-N1    & 1.79                          \\ \hline
Ga2-N2    & 1.85                          & Al2-N2    & 1.81                          \\ \hline
\end{tabular}
\end{table}

\begin{figure}[htb!]
	\centering
	\includegraphics[width=\linewidth]{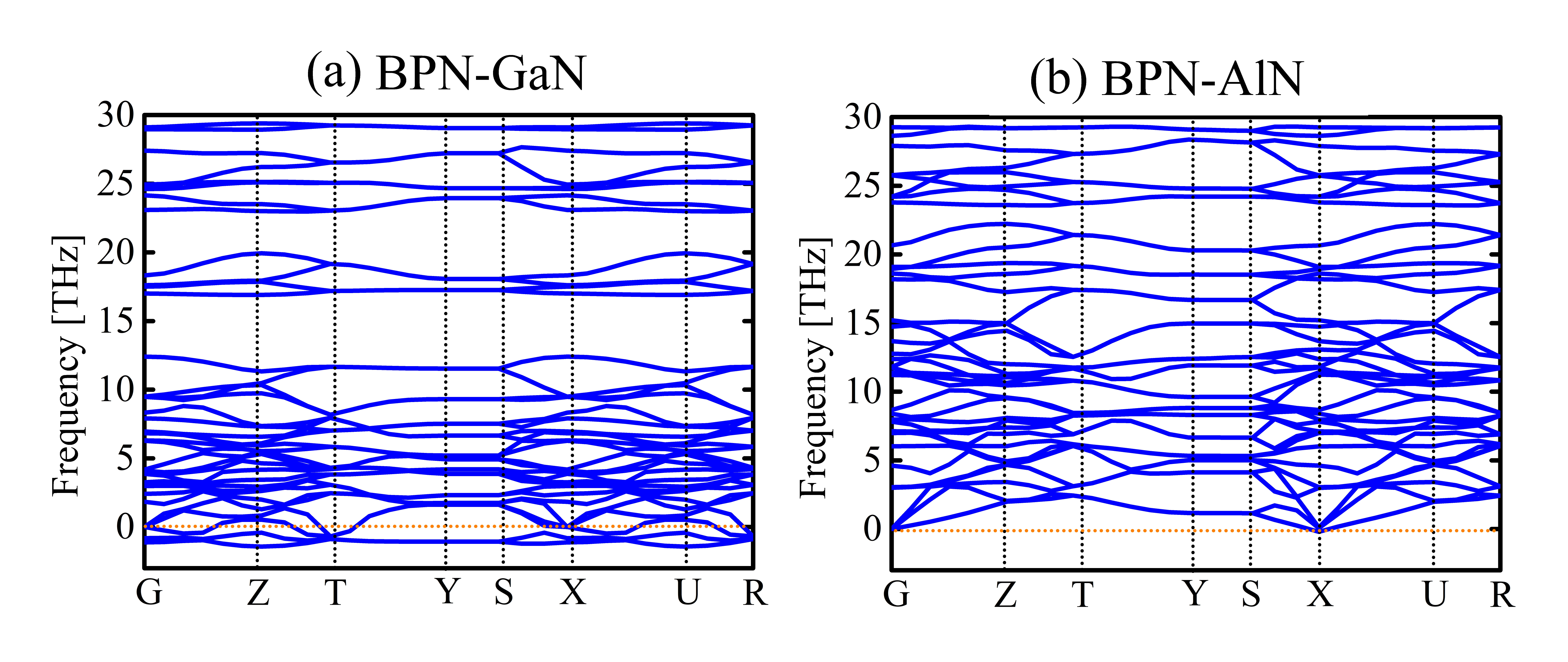}
	\caption{Phonon band structure of (a) BPN-GaN and (b) BPN-AlN calculated at GGA/PBE level.} 
	\label{fig:phonons}
\end{figure}

To assess the dynamical stability of the BPN-(Ga,Al)N sheets, we calculated their formation energies and the phonon dispersion relations along the high symmetry directions using the GGA/PBE method, as depicted in Figure \ref{fig:phonons}. We found formation energy values of -5.89 eV and -4.82 eV for BPN-AlN and BPN-GaN structures, respectively.

Phonon dispersion analysis of the unstrained BPN-GaN sheet (see Figure \ref{fig:phonons} (a)) revealed the presence of imaginary frequencies, specifically soft longitudinal and transverse acoustic modes \cite{Nika_2012,popovEPJB}. Generally, this trend indicates inherent instability. On the contrary, the phonon dispersion diagram shown in Figure \ref{fig:phonons}(b) consists exclusively of real frequencies, indicating the dynamic stability of the BPN-AlN sheets. The absence of a band gap between the acoustic and optical modes suggests a high scattering process rate and a short phonon lifetime, contributing to this material's relatively low lattice thermal conductivity.

The presence of imaginary frequencies in the BPN-GaN lattice indicates its susceptibility to structural distortions, emphasizing the significance of this observation. Such distortions may arise from inadequate bonding interactions within the material or even intrinsic stress. These instabilities can have adverse consequences, including phonon scattering, diminished thermal conductivity, and even structural deformations. The subsequent AIMD results below provide information on structural deformations of the BPN-GaN lattice.

One way to assess the thermal stability of a material using AIMD simulations is by monitoring the total energy per atom throughout the simulation. In a thermally stable system, the total energy per atom should remain relatively constant, indicating a balanced energy distribution within the material. Fluctuations in the total energy per atom during the simulation can provide insights into the material's response to thermal perturbations.

\begin{figure}[htb!]
	\centering
	\includegraphics[width=\linewidth]{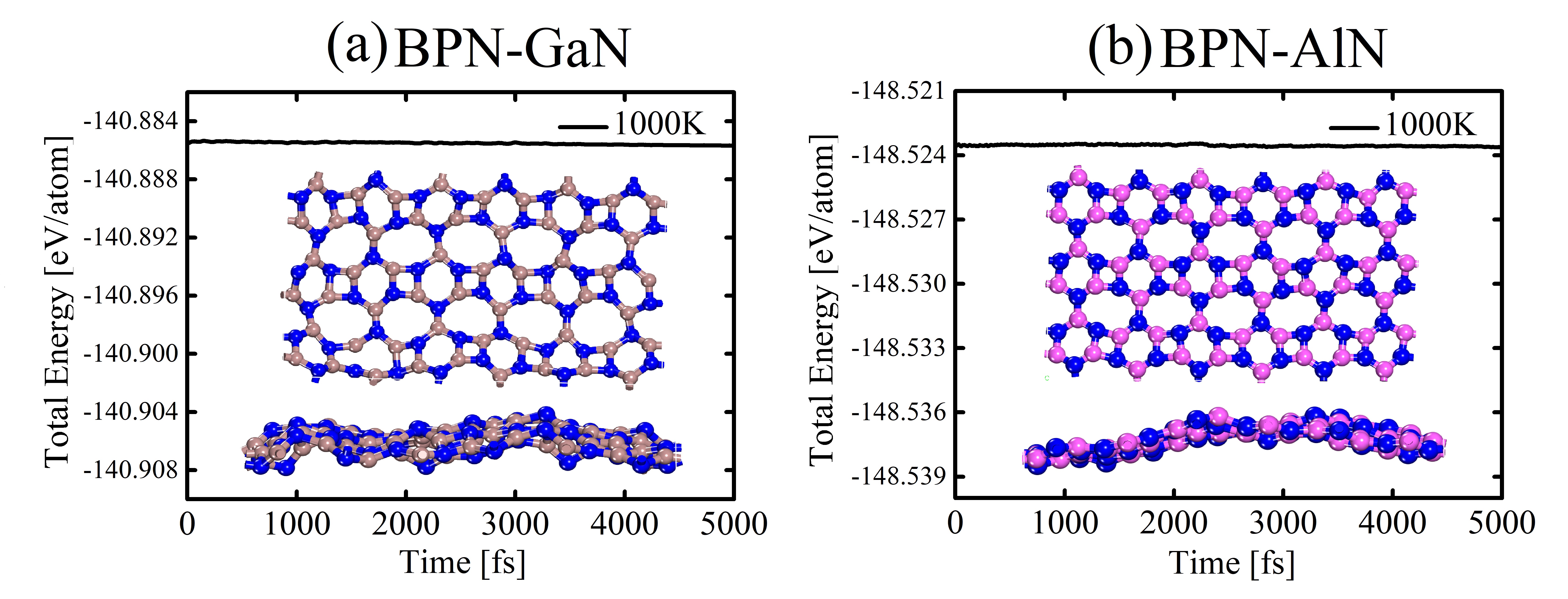}
	\caption{Time evolution of the total energy per atom in the (a) BPN-GaN and (b) BPN-AlN lattices at 1000K, using the GGA/PBE approach. The insets show the top and side views of the final AIMD snapshot at 5 ps.} 
\label{fig:AIMD}
\end{figure}

To ensure the dynamical stabilities of BPN-(Ga,Al)N, we conducted 5 ps of AIMD simulations at 1000K using the GGA/PBE approach, as depicted in Figure \ref{fig:AIMD}. The insets in the figure provide a top and side view of the AIMD snapshot, showcasing the lattices in their final stages during the simulations. A 3$\times$3$\times$1 supercell comprising 108 atoms was used. This period and system size are usually employed to investigate the system stability in AIMD simulations \cite{du2021cerium}. Observing both cases, one can see that the total energy per atom exhibits a flat pattern. The inset in Figure \ref{fig:AIMD}(a) presents the MD snapshot (side view) of BPN-GaN at 5 ps, revealing some lattice distortions imposed by the temperature. However, the top view illustrates that no bond reconstructions occur concerning the optimized lattice in Figure \ref{fig:sys}. Analogously, Figure \ref{fig:AIMD} (b) demonstrates that BPN-AlN maintains its planarity without any bond reconstructions. These observations suggest that both structures exhibit good thermal stability.   

\begin{figure}[htb!]
	\centering
	\includegraphics[width=0.6\linewidth]{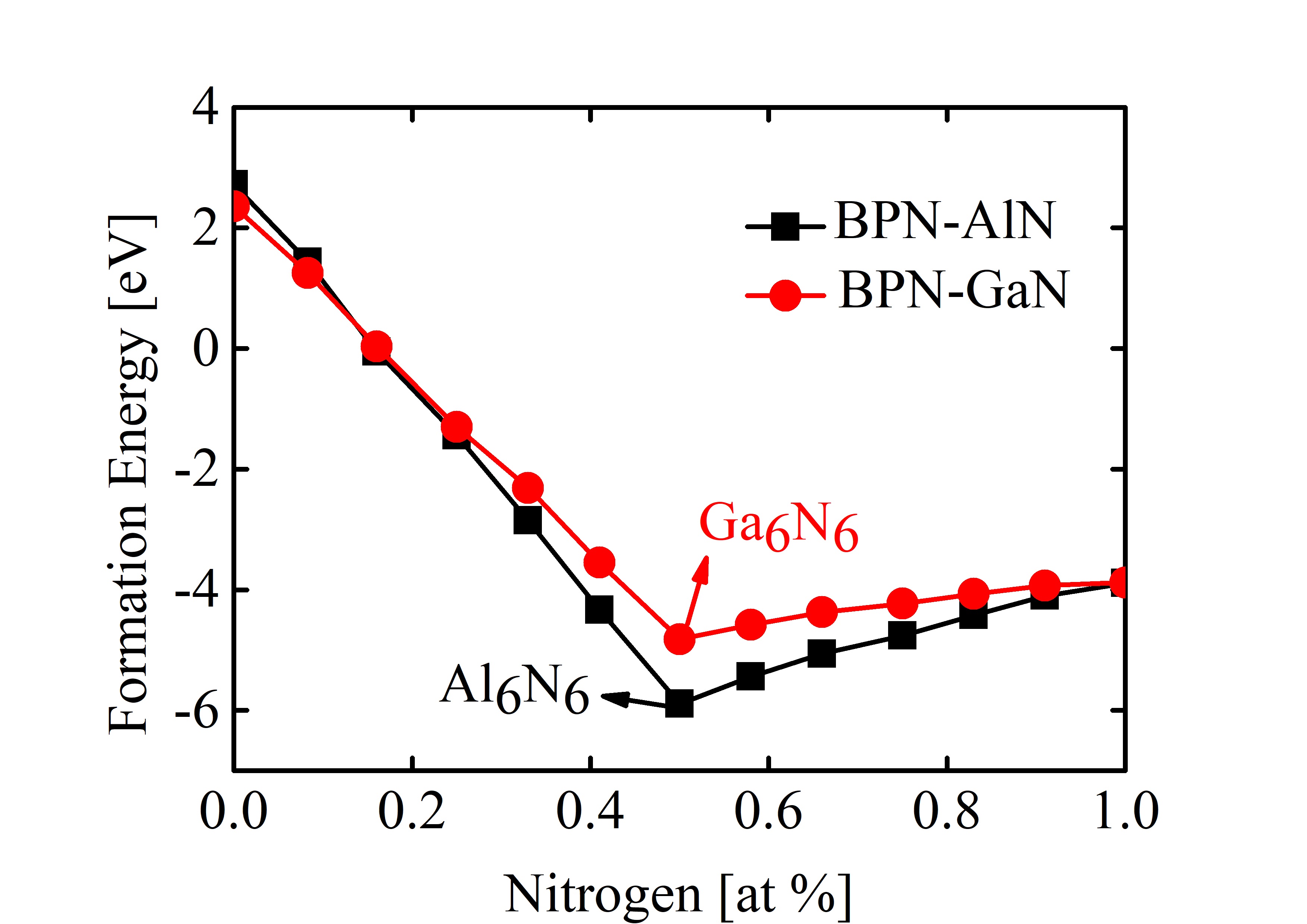}
	\caption{Computed formation energy as a function of the nitrogen concentration for BPN-like monolayers.} 
\label{fig:convexhull}
\end{figure}

When a structure's formation energy lies on the convex hull, it is considered thermodynamically stable and experimentally viable. Examining the formation energy plot versus N concentrations (refer to Figure \ref{fig:convexhull}), we have identified a global minimum structure in this 2D space. This minimum occurs precisely at the midpoint, with N:(Ga,Al) = 1:1 composition, resembling a structure closely akin to all-carbon BPN.

\begin{figure}[htb!]
	\centering
	\includegraphics[width=\linewidth]{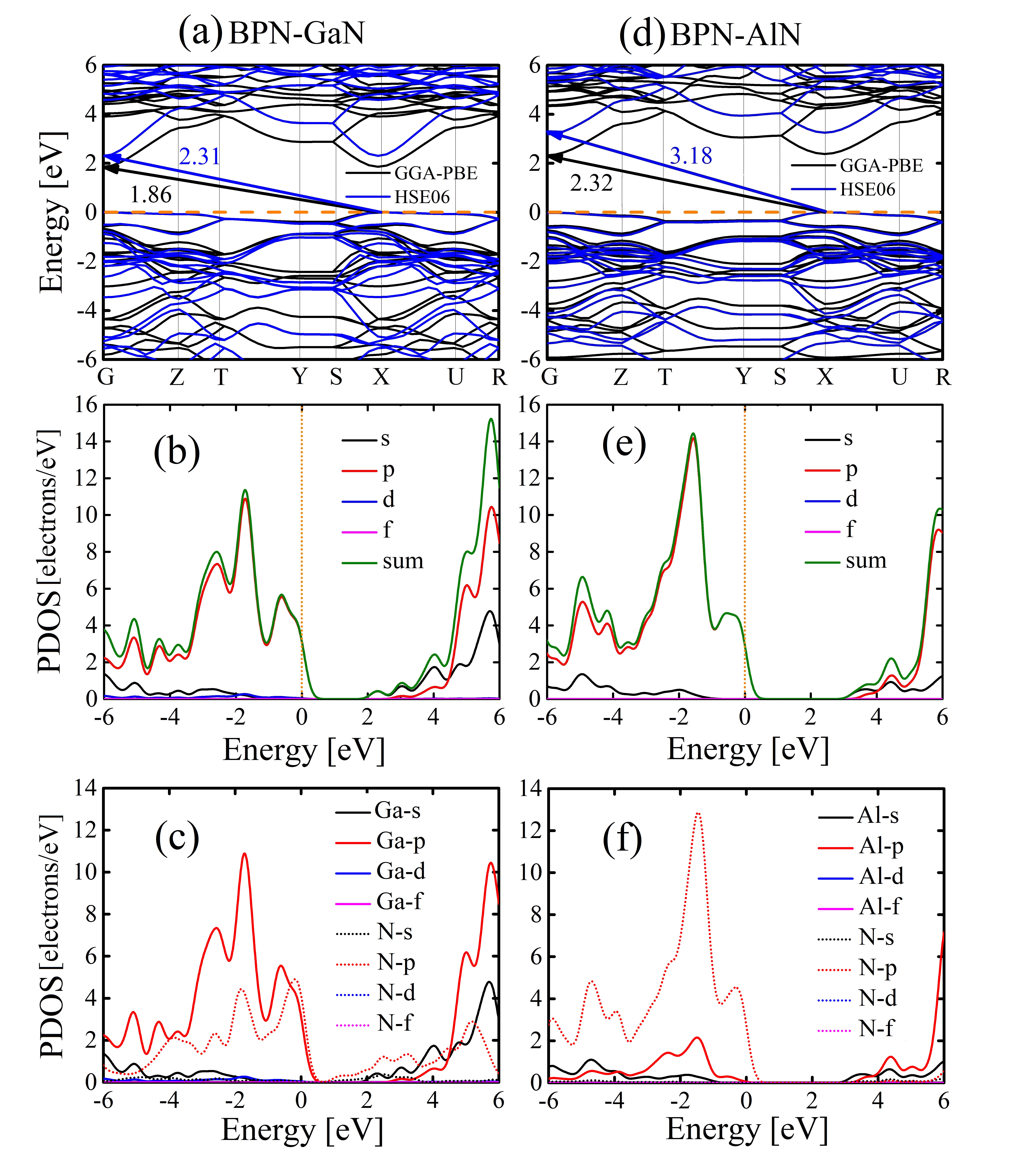}
	\caption{Electronic band structure for (a) BPN-GaN and (b) BPN-AlN sheets. These results were obtained using the GGA/PBE (black) and HSE06 (blue) approaches. Panels (b,e) and (c,f) show the total and per atom species partial density of states (PDOS) calculated at the HSE06 level, respectively.}
	\label{fig:bands}
\end{figure}

We now delve into the band structure characteristics of BPN-(Ga,Al)N lattices. The upper panels of Figure \ref{fig:bands} present the band structure profiles obtained using the GGA/PBE (black) and HSE06 (blue) methods. In the case of BPN-GaN (see Figure \ref{fig:bands}(a)), the band gap is indirect (from X-point to G-point) and measures 1.86 and 2.31 eV at the GGA/PBE and HSE06 levels, respectively. For BPN-AlN, the indirect band gap values from the X-point to the G-point are 2.32 eV and 3.18 eV according to the GGA/PBE and HSE06 calculations, respectively (see Figure \ref{fig:bands}(b)). 

Generally, GGA/PBE calculations underestimate band gaps \cite{10.1063/5.0059036}. Therefore, the hybrid functional HSE06 was used to obtain this work's electronic and optical properties precisely. The band structures of both BPN-(Ga,Al)N materials, as obtained from GGA/PBE, exhibit wide semiconducting band gaps, which are useful in optoelectronics and high-power devices \cite{10.1063/1.358463}. On the other hand, the HSE06 calculation suggests that BPN-AlN behaves as an insulator. These band structures are characterized by energy bands near the Fermi level that exhibit dispersion, indicating the delocalized nature of electronic states within these materials.

The difference in predicting the band gap values for BPN-AlN and BPN-GaN arises from these computational methods' inherent limitations and approximations. The HSE06 method, being a hybrid functional, incorporates a fraction of exact exchange in addition to the exchange-correlation functional. Including non-local exchange interactions tend to yield more accurate electronic structure predictions, especially for systems with significant electron localization or delocalization, like semiconductors. As such, the band gap calculated using HSE06 is often considered to provide a more reliable representation of electronic properties. In contrast, the PBE method relies on the generalized gradient approximation.

For BPN-AlN, the hole ($m_h^*$) and electron ($m_e^*$) effective masses are 0.250$m_0$ and 0.085$m_0$, respectively. For BPN-GaN, $m_h^*$ and $m_e^*$ are 0.875$m_0$ and 0.038$m_0$, respectively. $m_0$ is the electron mass. Generally, a lower effective mass correlates with higher carrier mobility, as carriers experience less resistance while moving in the crystal lattice. These values suggest that BPN-AlN could be particularly suitable for applications requiring efficient hole transport, such as p-type transistor channels or hole-dominant charge transport layers in photodetectors or solar cells. In contrast, BPN-GaN, with its higher hole effective mass and lower electron effective mass, is anticipated to have higher electron mobility, when contrasted with BPN-AlN, could be advantageous for applications that prioritize electron transport, such as n-type transistor channels or electron-dominant charge transport layers in optoelectronic devices. Remarkably, the $m_e^*$ values of BPN-(Al,Ga)N are lower than those for the typical \textit{wz}-AlN and \textit{wz}-GaN structures (0.22$m_0$ and 0.25$m_0$ \cite{persson2001effective}, respectively). This feature reinforces that BPN-(Al,Ga)N systems hold greater promise for applications demanding enhanced electronic transport efficiency than their wurtzite analogs.

We determined the full valence-band (VB) and conduction-band (CB) bandwidths for BPN-(Al,Ga)N systems. The VB and CB values for BPN-GaN are 17.76 eV and 25.70 eV, respectively. The VB and CB values for BPN-AlN are 16.29 eV and 21.94 eV, respectively. The bandwidth establishes a range of energy values over which electrons can exist within a particular band. It also captures the extent of energy values within which electrons can occupy states, thereby influencing their kinetic behavior and mobility. A wider energy bandwidth typically corresponds to a broader range of electron energies, which can impact various electronic phenomena, including conductivity, carrier mobility, and optical transitions. For instance, the VB values for common AlN and GaN monolayers are about 15.80 eV \cite{armenta2000ab}.

The partial density of states (PDOS) analysis of BPN-GaN and BPN-AlN, using the HSE06 hybrid functional, is shown in the left and right bottom panels of Figure \ref{fig:bands}, respectively. As a general trend, results indicate a significant predominance of p-states for both materials. In these cases, electronic transitions and interactions will primarily involve the participation of p-orbitals. These orbitals are typically associated with directional bonding. They are characterized by lobes with a dumbbell-like shape, enabling them to interact with neighboring atoms or orbitals in a controlled manner. 

Figures \ref{fig:bands}(b,e) provide a comprehensive view of the total contributions of s, p, d, and f orbitals for both atomic species within each lattice. Figures \ref{fig:bands}(c,f) illustrate the PDOS per atomic species. The p-states play a dominant role. In the case of BPN-GaN, Ga-p orbitals contribute significantly, followed closely by N-p states. Additionally, Ga-s orbitals primarily participate in forming the conduction band. For BPN-AlN, N-p orbitals have a higher contribution, with Al-p states following suit. Similarly, Al-s orbitals play a crucial role in shaping the conduction band.

Regarding the p-state properties, the behavior of p-states in a material holds crucial implications for its electronic, optical, and transport properties. A material's p-states can influence phenomena such as energy band gaps, absorption and emission of light, and charge mobility. For instance, the nature of p-states can determine the type of electronic transitions that occur, affecting the material's optical characteristics. Materials with well-defined p-states often exhibit interesting optoelectronic properties, making them suitable candidates for applications like photovoltaics, light-emitting diodes, and sensors. Furthermore, the interaction of p-states with neighboring atoms or external fields can influence charge transport mechanisms, affecting the material's performance in electronic devices.

Materials with well-defined p-states often exhibit interesting optoelectronic properties, making them suitable candidates for photovoltaics, light-emitting diodes, and sensors. Furthermore, the interaction of p-states with neighboring atoms or external fields can influence charge transport mechanisms, affecting the material's performance in electronic devices.

Here, it is worth comparing the BPN-AlN and BPN-GaN band structures with conventional AlN and GaN having wurtzite (\textit{wz}) and zincblende (\textit{zb}) structures \cite{zhang2020first}. Table \ref{tab:gaps} presents the theoretical band gap ($E_g$) using the HSE06+G$_0$W$_0$ approach and corresponding experimental (Exp) band gaps for (\textit{wz},\textit{zb})-GaN and (\textit{wz},\textit{zb})-AlN structures reported in the literature \cite{araujo2013electronic}. Notably, the BNP-AlN band gap significantly differs from those observed in (\textit{wz},\textit{zb})-AlN. While BPN-AlN acts as a wide-bandgap semiconductor (\textit{wz},\textit{zb})-AlN is an insulator. The band structure characteristics of BPN-AlN discussed above make it particularly suitable for emerging optoelectronic applications, such as high-frequency switching devices \cite{tazzoli2011ovenized}, light-emitting diodes \cite{taniyasu2006aluminium}, and RF power amplifiers \cite{piazza2012piezoelectric}. BPN-GaN displays similar band gap values to those reported for (\textit{wz},\textit{zb})-GaN. These materials also fall within the wide-bandgap semiconductors category, demonstrating the interesting potential for the aforementioned applications.

\begin{table}[htb!]
\caption{Comparison between theoretical band gap ($E_g$) using the HSE06+G$_0$W$_0$ approach and experimental band gaps ($Exp$), for AlN and GaN having wurtzite (\textit{wz}) and zincblende (\textit{zb}) structures. The values are given in electron-volt (eV), and they were reported in reference \cite{araujo2013electronic}.}
\centering
\label{tab:gaps}
\begin{tabular}{|c|c|c|}
\hline
Compound  & E$_{g}$ (HSE06+G$_{0}$W$_{0}$) [eV]& E$_{g}$ (Exp) [eV] \\    \hline
 wz-AlN   &      6.14     &   6.28                        \\     \hline
 wz-GaN   &      3.37     &   3.50                       \\     \hline
  zb-AlN  &      5.16     &   5.93                       \\    \hline
 zb-GaN   &      3.16     &   3.30                        \\    \hline
         
\end{tabular}
\end{table}

\begin{figure}[htb!]
	\centering
	\includegraphics[width=\linewidth]{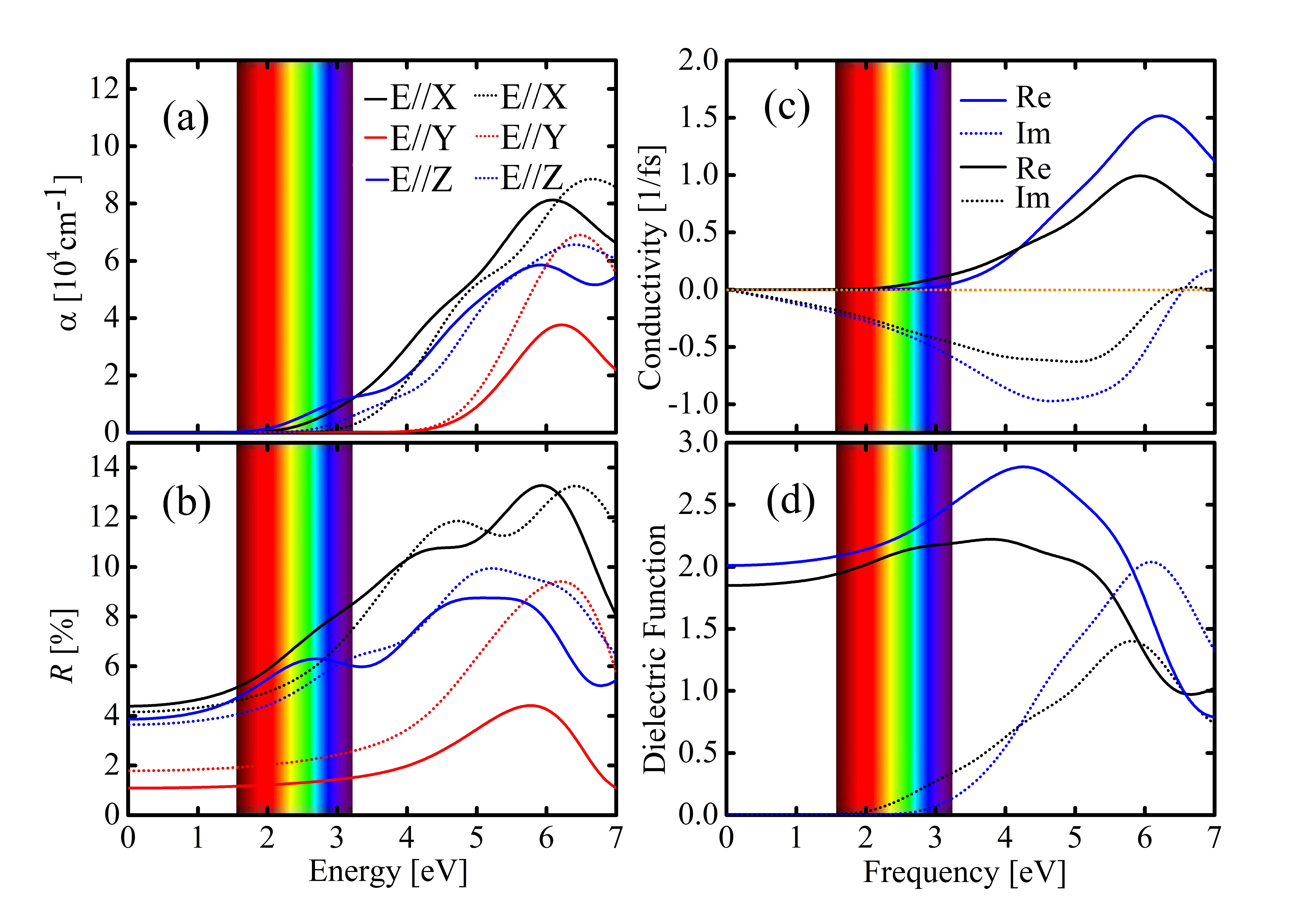}
	\caption{(a) absorption coefficient ($alpha$), (b) reflectivity ($R$), (c) optical conductivity ($\sigma$), (d) dielectric function for photon energies ranging from 0.0 to 7.0 eV, at HSE06 level, for polarised light beam oriented x (E//X), y (E//Y), and z (E//Z) directions related to the material's surface. The solid and dashed lines refer to the BPN-GaN and BPN-AlN systems.} 
	\label{fig:optical}
\end{figure}

Figure \ref{fig:optical} presents the optical properties of the BPN-(Ga, Al)N lattices, focusing on the polarization of light along the x (E//X), y (E//Y), and z (E//Z) directions. Calculations were performed using the HSE06 method. The solid and dashed lines correspond to the BPN-GaN and BPN-AlN systems. In Figure \ref{fig:optical}(a), the optical absorption ($\alpha$) is shown as a function of the photon energy. Both BPN-(Ga,Al)N materials exhibit strong optical activity, primarily in the ultraviolet (UV) region. 

The refractive index ($R$) in Figure \ref{fig:optical} (b) shows a prominent peak at approximately 6.0 eV for E//(X,Y) and 5.2 eV for E//Z in the case of BPN-GaN, while BPN-AlN shows peaks around 6.4 eV for E//(X,Y) and 5.5 eV for E//Z. The absorption coefficient gradually increases from zero and exhibits pronounced peaks at approximately 6.0 eV for E//(X,Z) and 6.2 eV for E//(Y) in the case of BPN-GaN, while BPN-AlN shows a peak at around 6.5 eV for E//(X,Y,Z). These findings suggest that BPN-(Ga,Al)N materials will likely be transparent in a solution and hold promise as potential candidates for UV device collectors.

In Figure \ref{fig:optical}(c), the real part (Re) of the optical conductivity is shown for BPN-GaN (solid black line) and BPN-AlN (solid blue line) over a photon energy range of 0.0 to 7.0 eV. Both materials exhibit distinct peaks in their curves, with a BPN-GaN peak at 5.8 eV and a BPN-AlN peak at 6.1 eV. The gradual increase in the real part of optical conductivity up to 5.8 eV corresponds to the increasing absorption coefficient within this energy range. These peaks indicate regions where electromagnetic waves can penetrate profoundly and reflect high optical conductivity. Therefore, targeting these specific peak energy values would be beneficial in optimizing BPN-(Ga,Al)N lattices for high optical conductivity.

In this figure, the imaginary part (Im) of the optical conductivity is displayed for BPN-GaN (dashed black line) and BPN-AlN (dashed blue line) over a photon energy range of 0.0 to 7.0 eV. Both materials exhibit peaks in their respective curves, with a BPN-GaN peak at 5.0 eV and a BPN-AlN peak at 4.7 eV. The negative values of the imaginary part of optical conductivity indicate an increase in the extinction coefficient, which suggests a reduction in the conductivity of the BPN-(Ga,Al)N lattices within this energy range. Consequently, the propagation of electromagnetic waves within this range is also diminished.

Figure \ref{fig:optical}(d) illustrates the real and imaginary parts of the complex dielectric function for BPN-(GaN,AlN) as a function of energy. In the energy range of 4.0-5.0 eV, both lattices exhibit peaks in their real parts, which can be attributed to the electronic polarizability of these materials. On the contrary, the imaginary part shows peaks at 6.0 eV, indicating electronic transitions involving the (Ga,Al)-p character of the conduction states, as revealed by the PDOS analysis in Figure \ref{fig:bands}. Beyond the band gap energy, the dielectric function shows a significant increase in its imaginary part, highlighting the strong UV absorption capabilities of the materials.

\begin{figure}[htb!]
	\centering
	\includegraphics[width=\linewidth]{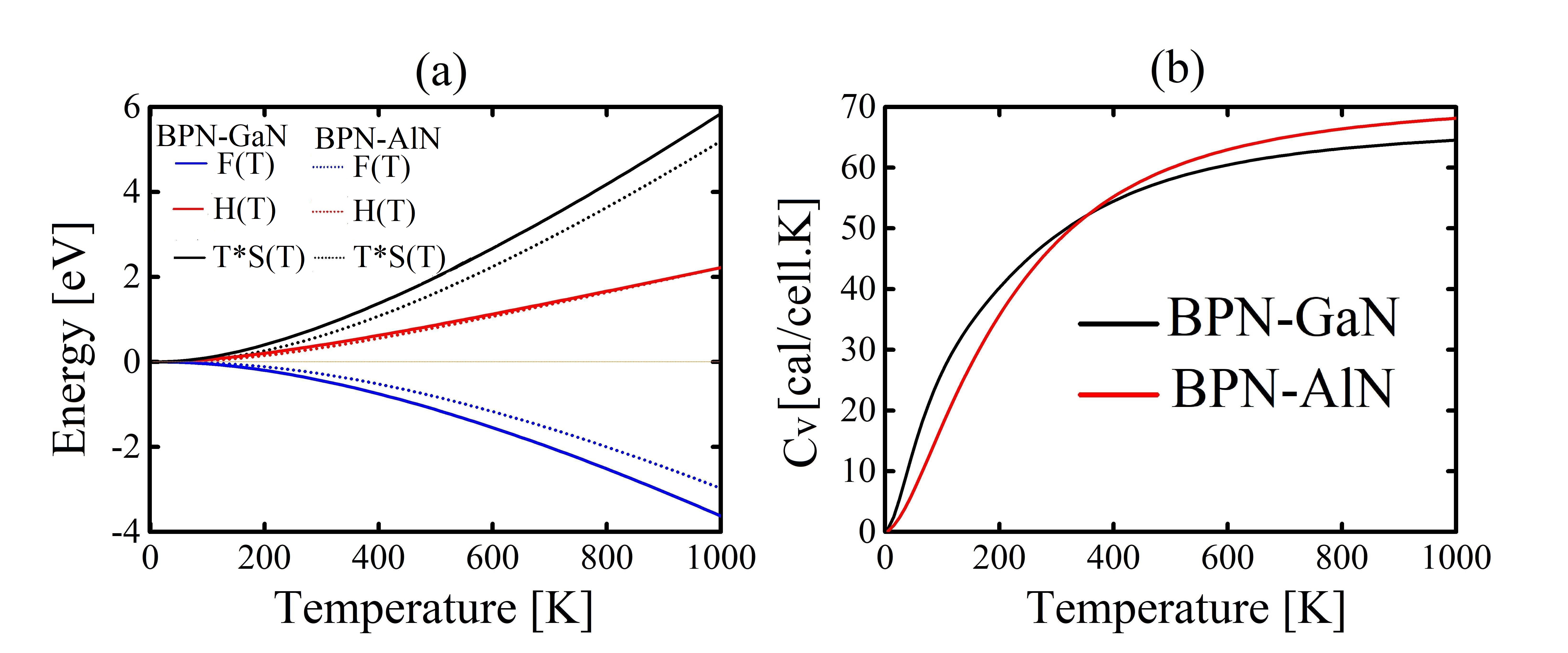}
	\caption{Thermodynamic potentials of BPN-(Ga,Al)N lattices. In panel (a), the red curves show the structure formation enthalpy $H(T)$, the blue curves represent the free energy $F(T)$, and the black curves illustrate the temperature times the entropy term ($T$*$S(T)$). In panel (b), the heat capacity as a function of the temperature ($C_V(T)$) is shown in black and red for BPN-GaN and BPN-AlN.}
	\label{fig:thermo}
\end{figure}

To gain insights into the thermal behavior of BPN-(Ga,Al)N lattices, we conducted calculations to determine their thermodynamic properties, including entropy $S(T)$, enthalpy $H(T)$, and free energy $F(T)$. We investigate the temperature dependence of the entropy term, expressed as $S(T) = U - F$, where $U$ represents the system's internal energy. Moreover, we evaluated the heat capacity $C_V(T)$ of the materials as a function of temperature. Figure \ref{fig:thermo} displays these calculated thermodynamic properties, which were obtained using the GGA/PBE level of theory.

Entropy measures the degree of disorder within a physical system's atomic arrangements. Figure \ref{fig:thermo} illustrates the temperature-dependent behavior of $T$*$S(T)$ in BPN-(Ga,Al)N lattices. At extremely low temperatures (below 100 K), the entropy value tends to zero, indicating a highly organized lattice structure with restricted atomic mobility. However, as the temperature rises, these monolayers exhibit increased thermal mobility, leading to a corresponding increase in the constituent particles' internal energy ($U$), including their kinetic energy. Consequently, the spacing between adjacent atoms (bond distances) expands, resulting in the crystals' thermal expansion. This increased particle instability correlates with an increase in entropy as the temperature continues to rise (as observed in the black curves of Figure \ref{fig:thermo}(a)).

Figure \ref{fig:thermo}(a) also presents the temperature dependence of $H(T)$ for the BPN-(Ga, Al) N monolayers. At low temperatures (below 100 K), $H(T)$ remains constant, indicating a stable state for both materials. However, a quadratic relationship between $H$ and $T$ is observed as the temperature exceeds this critical value. Lower values of $H(T)$ indicate a more favorable formation process, suggesting the thermodynamic stability of the lattice. This feature is consistent with the observation that higher temperatures result in higher $H(T)$, reducing the probability of finding an energetically stable structure, as depicted in Figure \ref{fig:thermo}(a). Notably, both materials exhibit $H(T)$ values below 6 eV for temperatures up to 1000 K.

It is essential to note the close relationship between $H(T)$ and $U$ in crystals. The internal energy ($U$) of the system is the sum of $H(T)$ and the product of pressure ($P$) and volume ($V$). As the temperature increases, the internal energy also increases. Consequently, the crystal undergoes thermal expansion, increasing volume ($V$) and pressure and volume product ($P\times V$). This trend, in turn, contributes to the elevation of $H(T)$ in BPN-(Ga,Al)N monolayers.

At low temperatures, lattices absorb thermal energy similarly to harmonic oscillators. However, as the energy reaches a certain threshold, lattice anharmonicity comes into play, resulting in a linear relationship. This behavior is observed in both systems, where the enthalpy increases with rising temperature, although the degree of dependence may vary. The system's internal energy ($U$ - $T$*$S$) is considered to understand this dependence further. Notably, the rate of increase in $T$*$S$ surpasses that of the internal energy $U$. Figure \ref{fig:thermo}(a) depicts the temperature dependence of $F(T)$. The lattices maintain their structural integrity at low temperatures without inducing anharmonicity or energy losses upon receiving additional energy. However, at high temperatures, lattice scattering becomes prominent, introducing anharmonicity into the system.

Heat capacity plays a vital role in assessing a material's thermal characteristics as it measures the energy absorption when the temperature increases by one Kelvin. In Figure \ref{fig:thermo}(b), the temperature-dependent behavior of the heat capacity in the BPN-(Ga, Al) N monolayers is depicted, revealing similarities between the two materials. BPN-AlN exhibits a slightly higher heat capacity, suggesting a greater production of phonons at a given temperature than that of BPN-GaN. As the temperature rises, the intrinsic monolayer heat capacity also increases due to the greater involvement of phonons. This trend aligns with the structural dependency observed in time-dependent enthalpy (see Figure \ref{fig:thermo}(a)).

In the temperature range of 0 to 300 K, the heat capacity experiences a steep increase with temperature, as shown in Figure \ref{fig:thermo} (b). However, between 300 and 1000 K, the growth rate slows down. The heat capacity reaches a maximum limit at temperatures beyond the Debye characteristic temperature (approximately 900 K for both materials), where all vibrations contain the same energy. This behavior follows the Dulong-Petit law, indicating that the specific heat capacity of the high-temperature lattice remains constant and independent of the temperature.

The temperature-dependent behavior of entropy and enthalpy sheds light on the stability and phase transitions of BPN-(Ga,Al)N monolayers. Constant enthalpy at low temperatures indicates a stable state, while the quadratic relationship observed as temperatures rise indicates thermodynamic stability. These findings have implications for the design of devices with enhanced thermal stability and controlled phase transitions. In this sense, BPN-(Ga,Al)N monolayers could serve as building blocks for smart materials capable of responding to specific temperature ranges and finding applications in thermal actuators or switches. The expansion of crystals' bond distances and the correlation between internal energy and enthalpy contribute to understanding thermal expansion mechanisms. These insights are crucial for developing nanoscale devices where precise control of thermal expansion is essential.

BPN-(Ga,Al)N structures could be integrated into nanoelectromechanical systems that require controlled thermal expansion for optimal performance. The temperature-dependent behavior of the heat capacity offers insights into energy absorption as temperature increases. Understanding heat capacity can guide the design of materials for efficient energy harvesting and conversion. BPN-(Ga,Al)N monolayers could be employed in thermoelectric generators, converting waste heat into usable electrical energy. The similarities and differences in heat capacity between BPN-AlN and BPN-GaN suggest distinct phonon behaviors that affect thermal conductivity. This insight is valuable for developing materials with tailored thermal transport properties. BPN-(Ga,Al)N monolayers could be engineered to enhance or minimize thermal conductivity, making them suitable for applications ranging from efficient thermal insulators to high-performance thermal conductors.

\begin{figure}[htb!]
	\centering
	\includegraphics[width=\linewidth]{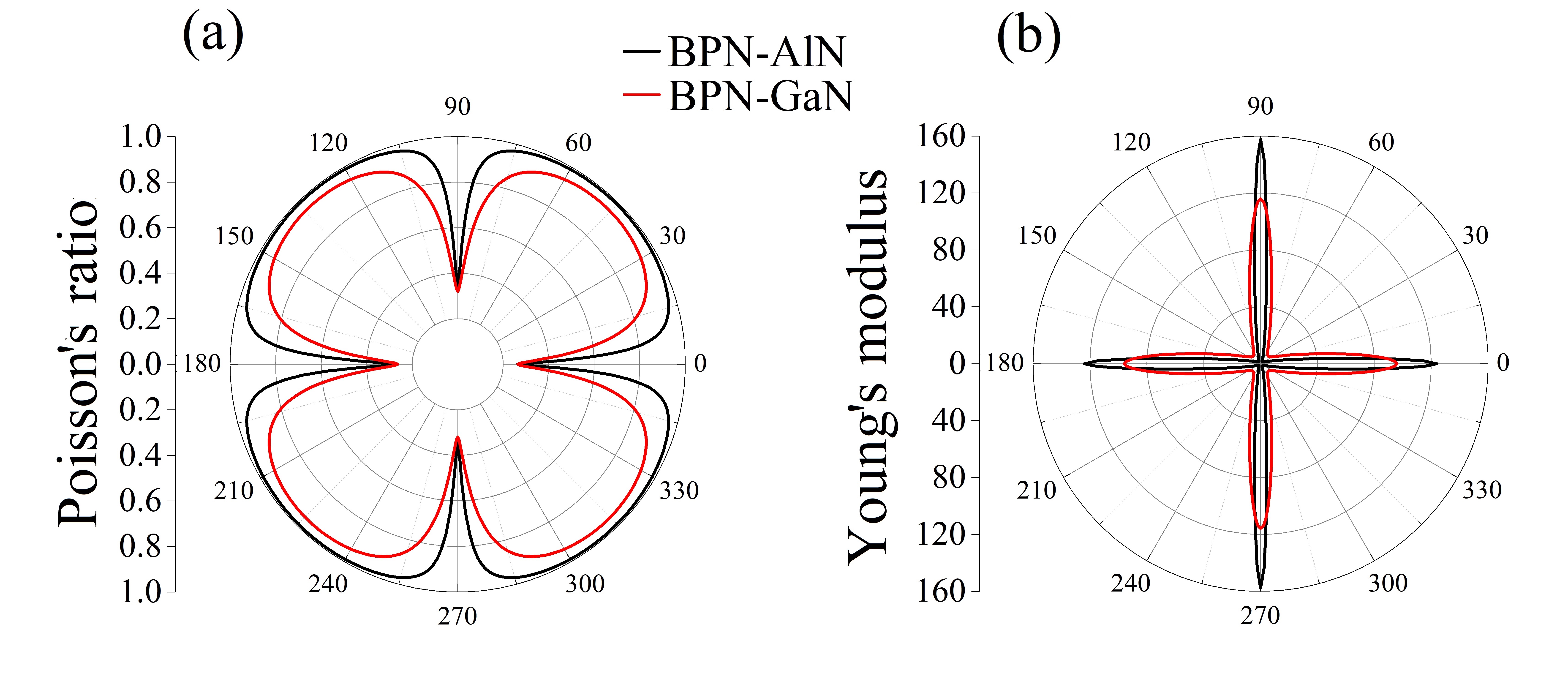}
	\caption{2D representation of (a) Poisson’s ratio and (b) Young's modulus in the xy plane for BPN-GaN in red and BPN-AlN in black.}
	\label{fig:elastic}
\end{figure}

We now analyze the elastic properties of BPN-(Ga,Al)N, which play a crucial role in understanding microcracks behavior and materials' overall durability. To assess the anisotropy in their mechanical properties, we determine the Poisson's ratio ($\nu(\theta)$) and Young's modulus ($Y(\theta)$) under pressure in the xy plane, as given by the following equations \cite{doi:10.1021/acsami.9b10472,doi:10.1021/acs.jpclett.8b00616}:

\begin{equation}
    \displaystyle Y(\theta) = \frac{{C_{11}C_{22} - C_{12}^2}}{{C_{11}\alpha^4 + C_{22}\beta^4 + \left(\frac{{C_{11}C_{22} - C_{12}^2}}{{C_{44}}} - 2.0C_{12}\right)\alpha^2\beta^2}}
    \label{young}
\end{equation}

\noindent and 

\begin{equation}
    \displaystyle \nu(\theta)= \frac{{(C_{11} + C_{22} - \frac{{C_{11}C_{22} - C_{12}^2}}{{C_{44}}})\alpha^2\beta^2 - C_{12}(\alpha^4 + \beta^4)}}{{C_{11}\alpha^4 + C_{22}\beta^4 + \left(\frac{{C_{11}C_{22} - C_{12}^2}}{{C_{44}}} - 2.0C_{12}\right)\alpha^2\beta^2}},
    \label{poisson}
\end{equation}

\noindent where, $\alpha=\cos(\theta)$ and $\beta=\sin(\theta)$. The elastic constants of BPN-(Ga,Al)N are shown in Table \ref{tab:elastic}. Furthermore, Figure \ref{fig:elastic} provides a 2D visualization of the Poisson's ratio (Figure \ref{fig:elastic}(a)) and Young's modulus (Figure \ref{fig:elastic}(b)) in the xy plane for BPN-(Ga,Al)N monolayers.

\begin{table}[htb!]
\centering
\caption{Elastic constants C$_{ij}$ (GPa) and maximum values for Young's modulus (GPa) ($Y_{MAX}$) and maximum ($\nu_{MAX}$) and ($\nu_{MIN}$) Poisson's ratios.}
\label{tab:elastic}
\begin{tabular}{| l |c|c|c|c|c|c|c|c|}
\hline
 Structure & C$_{11}$ & C$_{12}$ &C$_{22}$ &C$_{44}$ & $Y_{MAX}$  & $\nu_{MAX}$ & $\nu_{MIN}$ \\
 \hline
BPN-AlN    & $172.93$       & $45.44$     & $135.97$     & $0.46$  & $158.39$ & $0.91$ & $0.20$        \\
\hline
BPN-GaN    &$124.51$&$33.47$ & $104.73$ & $1.94$ &$114.50$ & $0.82$ & $0.11$                    \\
\hline
 \end{tabular}
\end{table}

The elastic constants C${11}$, C${22}$, C${12}$, and C${44}$ obtained for the BPN-(Ga,Al)N monolayers, presented in Table \ref{tab:elastic}, satisfy the Born-Huang criteria of the orthorhombic crystal ($C_{11}C_{22} - C_{12}^2>0$ and $C_{44}>0$) \cite{PhysRevB.90.224104,doi:10.1021/acs.jpcc.9b09593}, indicating their good mechanical stability. Additionally, Young's modulus and Poisson's ratio of BPN-(Ga,Al)N structures were calculated using Equations \ref{young} and \ref{poisson}, respectively (see Figure \ref{fig:elastic}). We found these properties exhibit anisotropic characteristics.

Regarding Young's modulus, the calculated maximum values ($Y_{MAX}$) are 158.39 GPa and 114.50 GPa for BPN-AlN and BPN-GaN, respectively. These values are smaller than those of hexagonal AlN (405.1 GPa \cite{doi:10.1021/acs.jpcc.6b09706}) and GaN (287.7 GPa \cite{10.1063/5.0063765}) monolayers. These differences can be attributed to the porosity of the BPN-like structures, resulting from the presence of eight-atom rings and the rigidity of the bonds in the four-atom rings. These factors contribute to the reduced resilience of BPN-(Ga,Al)N monolayers to strain compared to their hexagonal counterparts \cite{D1NR07959J}. 

When a compressive (tensile) strain is acted in one direction, materials tend to expand (contract) in the perpendicular direction, corresponding to the positive Poisson’s ratio for ordinary materials. The opposite is for materials with a negative Poisson's ratio (i.e. auxetic materials \cite{Mazaev_2020}). In Figure \ref{fig:elastic}, one can note that BPN-AlN and BPN-GaN have positive Poisson's ratio, with maximum values ($\nu_{MAX}$) about 0.91 and 0.82, respectively. For these $\nu_{MAX}$ values, Young's modulus values are less than 20 GPa, indicating the incompressibility of these materials for biaxial strains, i.e., strains that are not applied solely along the x or y directions (uniaxial strains).  

Most common materials exhibit Poisson's ratios ranging from 0.2 to 0.5 \cite{greaves2011poisson}. A Poisson's ratio of 0.5 indicates incompressible materials that retain their lateral dimensions when subjected to axial strain. For BPN-AlN and BPN-GaN, the minimum values of their Poisson's ratios ($\nu_{MIN}$) for uniaxial strains applied along the x and y directions are 0.20 and 0.11, respectively. These values are significantly lower than those observed in hexagonal AlN and GaN, which are approximately 0.46 and 0.44 \cite{Luo_2019}, respectively. The lower Poisson ratios of BPN-AlN and BPN-GaN can be attributed to their BPN-like structures, which possess a degree of porosity higher than those of their honeycomb-based lattice counterparts. This increased porosity allows BPN structures to undergo more deformation under tension, resulting in lower Poisson ratios.

Notably, BPN-AlN and BPN-GaN exhibit distinctive mechanical responses due to their unique BPN-like structures. Young's modulus values of 158.39 GPa and 114.50 GPa for BPN-AlN and BPN-GaN are lower than those for their hexagonal counterparts \cite{zhang2021dft,le2014atomistic}. This deviation is attributed to the porosity inherent in the BPN-like structures, which results from the presence of eight-atom rings and the rigidity of bonds in the four-atom rings. Consequently, these structural features influence the mechanical resilience of BPN-(Ga,Al)N monolayers, rendering them less resistant to strain than hexagonal AlN and GaN monolayers.

The anisotropic mechanical behavior, reflected in Young's modulus, makes BPN-AlN and BPN-GaN promising candidates for flexible electronics and nanomechanical devices. Their relatively lower Young's modulus than hexagonal counterparts suggests improved deformability and resilience to bending or mechanical stress. This attribute is crucial in applications such as wearable electronics, sensors, and flexible displays, where mechanical adaptability is a key consideration.

The Poisson's ratio of BPN-AlN and BPN-GaN offers an intriguing avenue for microelectromechanical systems applications. The positive Poisson's ratio indicates responsiveness to strain variations, while the lower values compared to hexagonal materials hint at unique deformability. These attributes can be leveraged to design MEMS devices with tailored actuation and sensing capabilities. Moreover, the distinctive mechanical properties of BPN-(Ga,Al)N monolayers could contribute to energy harvesting applications. Their good deformability under mechanical stress can be interesting for piezoelectric energy conversion.

Based on the results presented here, the advantages of BPN-(Al,Ga)N materials over common GaN and AlN with wurtzite and zincblende structures are twofold: I. \textit{Novel electronic properties:} BPN-based materials exhibit a unique atomic arrangement that can lead to innovative electronic properties, holding promise for electronic and optoelectronic devices; II. \textit{Porous structure and surface area:} The BPN topology boasts a substantial surface area-to-volume ratio and a porous composition, rendering it suitable for various surface-dependent applications, including gas capture, sensing, and adsorption \cite{chen2023new,al2022two,zhang2023novel}.

We must acknowledge two primary disadvantages of BPN-(Al,Ga)N materials compared to conventional GaN and AlN structures: I. \textit{Synthesis and stabilization challenges:} The synthesis and stabilization of BPN pose challenges due to its intricate structure and potential reactivity; II. \textit{Ongoing research and exploration:} BPN remains a relatively new material, with its properties and potential applications still actively researched.

Turning to the strengths of common AlN and GaN structures, the piezoelectric properties of \textit{wz}-AlN and \textit{wz}-GaN enable the efficient conversion between electrical and mechanical energy forms. These properties remain unexplored in BPN-based materials. Moreover, \textit{wz}-GaN finds extensive utility in LEDs \cite{guha1998ultraviolet}, laser diodes \cite{fujimoto2013visible}, and other optoelectronic devices owing to its direct band gap and high electron mobility. The zincblende structure facilitates the creation of quantum dots, holding promise for applications in quantum computing and advanced optical devices. Nonetheless, challenges in crystal growth persist, impacting the achievement of high-quality single crystals and subsequently affecting device performance. The synthesis and control of zincblende structures also present challenges, influencing device integration and overall performance.

\begin{figure}[htb!]
	\centering
	\includegraphics[width=\linewidth]{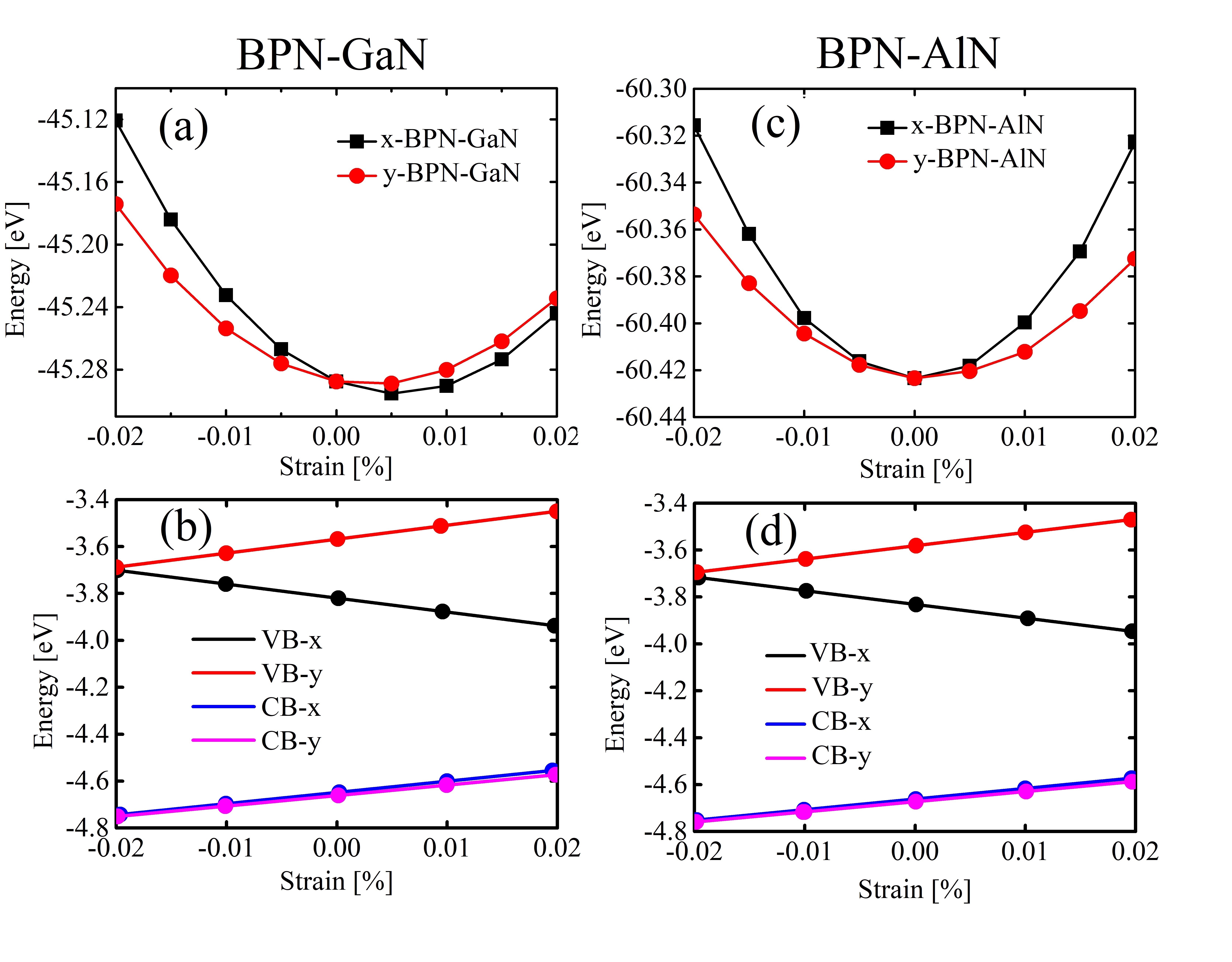}
	\caption{Panels (a,b) illustrate the quadratic fitting curve for the BPN-(Ga,Al)N total energy as a function of the applied compressive and tensile strains to derive the in-plane stiffness ($C_{2D}$). Panels (c,d) show the linear fitting diagram for the BPN-(Ga,Al)N total energy as a function of the compressive and tensile strains to derive the deformation potential constant ($E_1$). } 
\label{fit}
\end{figure}

Most of the notable properties of these materials derive from their band gap configurations. With band gap values of 2.30 eV for BPN-AlN and 3.18 eV for BPN-GaN, these materials offer a tunable range of band gaps, enabling tailored electronic and optical functionalities. The variation in band gaps allows for design flexibility in different devices and applications. Moreover, BPN-AlN and BPN-GaN exhibit wide band gaps, making them suitable for high-power and high-temperature environments.

When contrasted with conventional AlN structures, the comparatively lower band gap of BPN-AlN could make it suitable for developing high-efficiency light-emitting diodes, particularly in the UV-visible spectral range. The wider BPN-GaN band gap lends itself to optoelectronic devices operating at higher photon energies. BPN-AlN and BPN-GaN, with their wide band gap and potential for efficient electron mobility, hold promise for high-frequency switching applications, such as RF power amplifiers and high-speed transistors.

Leveraging their unique electronic properties, BPN-AlN and BPN-GaN can find utility in sensor technologies, including gas sensing, biosensing, and radiation detection, where their porous lattice structure and tailored band gaps could enhance sensitivity and selectivity. The tunable band gaps of these materials suggest applications in photovoltaic devices, where they could facilitate efficient light absorption and electron-hole separation.

To comprehensively understand the electronic conductance in monolayer BPN-(Ga,Al)N, we have extended our analysis to calculate charge carrier mobility using the deformation potential (DP) theory \cite{bardeen1950deformation}. This approach is based on the effective mass approximation and is commonly employed for evaluating carrier mobility in 2D layered semiconductors. Carrier mobility ($\mu$) is defined as

\begin{equation}
\displaystyle \mu=\frac{e\hbar C_{2D}}{k_BTm^{*}_{i}m_{d}\left(E_1\right)^2},
\end{equation}

\noindent where $m_d=\sqrt{m_x^*m_y^*}$ represents the average effective mass. $C_{2D}$ is the in-plane stiffness $C_{2D}=\left(\partial^2E/\partial\delta^2\right)/S_0$ in which $E$, $\delta$, and $S_0$ are the total energy, applied strain, and the area of the investigated system, respectively. $m_{i}^{*}$ denotes the effective mass for electrons ($e$) and holes ($h$) according to Equation \ref{mass}. $E_1=dE_{edge}/d\delta$, where $\delta$ is the applied strain by a step of 0.5\% and $E_{edge}$ is the energy of the band edges (VBM for the holes and CBM for the electrons). VBM and CBM stand for valence band maximum and conduction band minimum, respectively, located at G and X points. $k_B$, $T$, $e$, and $\hbar$, correspond to Boltzmann's constant, temperature, the elementary charge of an electron, and Planck's constant. 

\begin{table}[htb!]
\label{tab2}
\begin{tabular}{|c|c|c|c|c|c|c|}
\hline
System                                                                                               & Carrier    & $m_i^*$ ($m_0$) & $m_d$ ($m_0$) & $C_{2D}$ (eV/\r{A}) & $E_1$ (eV) & $\mu$ (10$^3$cm$^2$V$^{-1}$s$^{-1}$) \\ \hline
\multicolumn{1}{|l|}{\multirow{2}{*}{\begin{tabular}[c]{@{}l@{}}BPN-GaN\\ x-direction\end{tabular}}} & \textit{e} & 0.45     & 0.45    & 9.21       & 4.78    & 28.43          \\ \cline{2-7} 
\multicolumn{1}{|l|}{}                                                                               & \textit{h} & 0.59     & 0.76    & 9.21       & -5.95   & 10.70          \\ \hline
\multicolumn{1}{|l|}{\multirow{2}{*}{\begin{tabular}[c]{@{}l@{}}BPN-GaN\\ y-direction\end{tabular}}} & \textit{e} & 0.46     & 0.45    & 7.40       & 4.46    & 25.67          \\ \cline{2-7} 
\multicolumn{1}{|l|}{}                                                                               & \textit{h} & 0.98     & 0.76    & 7.40       & 5.96    & 5.19           \\ \hline
\multirow{2}{*}{\begin{tabular}[c]{@{}c@{}}BPN-AlN\\ x-direction\end{tabular}}                       & \textit{e} & 0.23     & 0.58    & 9.67       & 4.52    & 57.54          \\ \cline{2-7} 
                                                                                                     & \textit{h} & 1.80     & 2.44    & 5.57       & -5.80   & 1.25           \\ \hline
\multirow{2}{*}{\begin{tabular}[c]{@{}c@{}}BPN-AlN\\ y-direction\end{tabular}}                       & \textit{e} & 1.45     & 0.58    & 9.67       & 4.33    & 9.94           \\ \cline{2-7} 
                                                                                                     & \textit{h} & 3.31     & 2.44    & 5.57       & 5.69    & 0.71           \\ \hline
\end{tabular}
\caption{Calculated effective mass $m_i^*$, average effective mass $m_d$, in-plane stiffness $C_{2D}$, DP constant $E_1$, and charge carrier mobility $\mu$ for BPN-(Ga.Al)N monolayer. Here, $m_0$ is the effective mass of a free electron.}
\end{table}

We have summarized the computed values for $m_{i}^{*}$, $m_d$, $C_{2D}$, $E_1$, and $\mu$ at 300 K in Table \ref{tab2}. By subjecting the system to both compressive and tensile strains, we calculate $E_1$ through linear fitting of CBM and VBM values for electrons and holes, respectively (see Figure \ref{fit}(a)). Subsequently, $C_{2D}$ is determined by applying a quadratic fitting approach to the total energy data as a function of compressive and tensile strains, as depicted in Figure \ref{fit}(b).

The electron mobility surpasses the hole mobility, as indicated in Table \ref{tab2}. Furthermore, the carrier mobility in the BPN-(Ga,Al)N monolayer exhibits anisotropic behavior, with notable distinctions between the x and y directions. Specifically, the carrier mobility along the x direction is considerably higher than the y direction. This discrepancy is primarily attributed to the significantly smaller effective mass in the x direction than in the y direction.

For instance, in h-AlN lattices, the electron mobility is 0.32 and 0.295 (in units of 10$^3$ cm$^2$V$^{-1}$s$^{-1}$) in the x and y directions, respectively \cite{ren2020high}. In contrast, the hole mobility measures 1.77 and 1.8 (in units of 10$^3$ cm$^2$V$^{-1}$s$^{-1}$) in these directions  \cite{ren2020high}. Notably, both electrons and holes in BPN-AlN exhibit higher mobility than h-AlN. Similarly, in h-GaN lattices, the electron mobility is 0.29 and 0.30 (in units of 10$^3$ cm$^2$V$^{-1}$s$^{-1}$) in the x and y directions, respectively, while the hole mobility is 1.74 and 2.39 (in units of 10$^3$ cm$^2$V$^{-1}$s$^{-1}$) in the respective directions  \cite{ren2020high}. In BPN-GaN, both electrons and holes exhibit enhanced mobility compared to h-GaN. The superior charge carrier mobility observed in the BPN-(Ga,Al)N systems compared to h-(Ga,Al)N can be attributed to their distinct topologies.

\section{Conclusions}

In summary, we performed DFT calculations to study the mechanical, electronic, thermodynamic, and optical properties of two group-III counterparts of the recently synthesized biphenylene network (BPN): gallium nitride (BPN-GaN) and aluminum nitride (BPN-AlN). We calculated formation energies and phonon dispersion relations along the high symmetry directions to assess their dynamical stabilities. Moreover, AIMD simulations were performed to describe their dynamical stability at 1000 K.

The phonon dispersion profile of the BPN-GaN sheet revealed the presence of imaginary frequencies. Conversely, the phonon dispersion diagram of the BPN-AlN monolayer exclusively consists of real frequencies, indicating its dynamic stability. The final AIMD snapshot of BPN-(Al,Ga)N revealed lattice distortions. However, no bond reconstructions occur concerning the initially optimized lattices. These observations suggest that both structures exhibit good thermal stability.

In the case of BPN-GaN, the band gap is indirect and measures 2.31 eV and 1.86 eV at the HSE06 and GGA/PBE levels, respectively. For BPN-AlN, the indirect bandgap values are 2.32 eV and 3.18 eV, according to GGA/PBE and HSE06 calculations, respectively. The band structures of both BPN-(Ga,Al)N materials exhibit wide semiconducting band gaps, which find applications in optoelectronics and high-power devices.  On the other hand, the HSE06 calculation suggests that BPN-AlN behaves as an insulator. The optical properties analysis demonstrated strong optical activity in the ultraviolet region for both BPN-(Ga,Al)N materials, making them potential candidates for UV device collectors.

\section*{Acknowledgement}

This work was financed by the Coordenação de Aperfeiçoamento de Pessoal de Nível Superior (CAPES) - Finance Code 001, Conselho Nacional de Desenvolvimento Cientifico e Tecnológico (CNPq), and FAP-DF. L.A.R.J acknowledges the financial support from FAP-DF grants $00193-00000853/2021-28$, $00193-00000857/2021-14$, and $00193-00000811/2021-97$, FAPDF-PRONEM grant $00193.00001247/2021-20$, and CNPq grants $302236/2018-0$ and $350176/2022-1$. L.A.R.J. acknowledges N\'ucleo de Computaç\~ao de Alto Desempenho (NACAD) and for providing the computational facilities. This work used resources of the Centro Nacional de Processamento de Alto Desempenho em São Paulo (CENAPAD-SP). 

\bibliographystyle{unsrt}
\bibliography{bibliography.bib}
	
\end{document}